\documentstyle[twocolumn,prb,aps,epsfig,floats]{revtex}
\begin{document}
\twocolumn[\hsize\textwidth\columnwidth\hsize\csname 
@twocolumnfalse\endcsname
\title{Anisotropic random resistor networks: a model for 
piezoresistive response of thick-film resistors. } 
\author{C. Grimaldi$^1$, P. Ryser$^1$, and S. Str\"assler$^{1,2}$} 
\address{$^1$ Institut de Production et Robotique, LPM,
Ecole Polytechnique F\'ed\'erale de Lausanne,
CH-1015 Lausanne, Switzerland}
\address{$^2$ Sensile Technologies SA, PSE, CH-1015 Lausanne, Switzerland}

\maketitle

\centerline \\

\begin{abstract}
A number of evidences suggests that thick-film resistors
are close to a metal-insulator transition and that tunneling
processes between metallic grains are the main source of
resistance. We consider as a minimal model for description of
transport properties in thick-film resistors a percolative
resistor network, with conducting elements governed by tunneling.
For both oriented and randomly oriented networks, we show that
the piezoresistive response to an applied strain is model dependent
when the system is far away from the percolation thresold, while
in the critical region it acquires universal properties. In particular
close to the metal-insulator transition, the piezoresistive anisotropy
show a power law behavior. Within this region, there exists a simple 
and universal relation between the conductance and the piezoresistive 
anisotropy, which could be experimentally tested by common cantilever
bar measurements of thick-film resistors. 

\end{abstract}
\vskip 2pc ] 

\narrowtext

\section{Introduction}
\label{intro}
Thick-film resistors (TFRs) are a particular class of granular metals
made of RuO$_2$, Bi$_2$Ru$_2$O$_7$, or other metal-oxide granules
embedded in an insulating glassy matrix. The importance of TFRs in applicative
problems is due to their large piezoresistive response which, together with
high stability and low manufacturing costs, makes the TFRs highly competitive
as pressure and force sensors.\cite{white,prude1} Hence, the full understanding of the factors 
governing the piezoresistive response is a very important issue 
in optimizing the performances of TFRs. 
The interest in the piezoresistive response however
concerns also more fundamental aspects connected, as we discuss in the following,
with the role of transport anisotropy in disordered systems.

In practical measurements, the piezoresistive response is obtained by
the so-called gauge factor $K$, commonly defined as $K=\delta R/\varepsilon R$,
where $\delta R$ is the change of resistance $R$ for a given applied strain 
$\varepsilon$. When the strain is directed along the voltage drop,
the change in resistance determines the longitudinal gauge factor $K_{\rm L}$.
Usually, in TFRs the gauge factor ranges from $K_{\rm L}\sim 2$
up to $K_{\rm L}\sim 35$ or more,\cite{prude1} following roughly the 
empirical relation\cite{prude2}
\begin{equation}
\label{empi}
K_{\rm L}\sim \ln R.
\end{equation}
Both the high values of $K_{\rm L}$ and Eq.(\ref{empi}) point towards
a scenario in which tunneling processes between metallic grains 
are the main source of resistance.\cite{prude2,pike}
In fact, if $d$ is the mean tunneling distance, then transport between adjacent 
metallic grains is approximatively
\begin{equation}
\label{tunnel1}
R\propto \exp(2d/\xi),
\end{equation}
where $\xi$ is the localization length. Let us assume that under an
applied strain $\varepsilon$ the distance $d$ changes to $d\rightarrow
d+\delta d=d(1+\varepsilon)$, where we have set $\varepsilon=\delta d/d$. 
The strain-induced change of $R$ is therefore
$\delta R/R\sim 2d\varepsilon/\xi$, which implies the empirical
relation Eq.(\ref{empi}).\cite{prude2}

In obtaining this simple result, we have assumed that the direction of
intergrain tunneling is along the direction of the applied strain.
In this way we obtained an upper limit of the longitudinal gauge factor $K_{\rm L}$.
In structurally isotropic systems, however, the direction between two 
neighbouring grains is oriented randomly and can be even ortogonal to the
direction of $\varepsilon$. This leads to a reduction of $K_{\rm L}$
and to a contemporary appearance of a nonzero transverse gauge factor $K_{\rm T}$
defined by the change of resistance when the voltage drop is ortogonal to
the direction of $\varepsilon$. In addition, if we consider also that TFRs are
disordered conductors, then it is natural to expect that, in flowing
from one end to the other of the sample, the microscopic currents are 
subjected to drastic changes of direction sampling therefore regions where
the intergrain distance is $d(1+\varepsilon)$ and regions where $d$ 
remains unstrained. This effect contributes to the decrease of $K_{\rm L}$ and 
to the enhancement of $K_{\rm T}$.\cite{grima1}

What is commonly observed in TFRs is that $K_{\rm L}$ and $K_{\rm T}$ are 
comparable,\cite{prude2,hrovat}
indicating therefore that the current tortuosity induced by disorder and random
bond orientations is an important element and needs further investigation.
To this end, we have recently addressed the problem of the piezoresistance in a
simplified model of TFRs (or more generally of granular metals) consisting of
a random resistor network with a fraction $p$ of bonds having resistances of
the form of Eq.(\ref{tunnel1}) and the remaining fraction $1-p$ having zero
conductance.\cite{grima1} In accord with the qualitative discussion of above, we have
found that as $p$ is reduced from the unity, the longitudinal gauge 
factor is diminished and the transverse one is enhanced. 
This trend holds true all the way down to the
critical bond concentration $p_c$ at which the conductance $G$ vanishes and 
$K_{\rm L}\rightarrow K_{\rm T}$ for an uniaxial strain.\cite{grima1} 
The observation that $K_{\rm L}\sim K_{\rm T}$
in TFRs could therefore be an indication that these systems are close 
to a metal-insulator transition, an hypothesis sustained by the very low
metallic volume fractions $v$ in TFRs and the observation that the conductance
$G$ follows a power law behavior $G\sim (v-v_c)^t$, where $v_c$ is the
critical volume fraction and $t$ is a critical exponent.\cite{carcia,szpytma,bobran,kusy}

The aim of this paper is to study further the problem of piezoresistive 
response in random resistor networks by focusing on the equivalence between
transport in anisotropic networks and the piezoresistive response of
isotropic ones. Close to the percolation thresold, the two systems are equivalent
in the sense that transport is governed by the same critical exponents.
In addition we consider also the effect of random bond orientations and provide
explicit formulas from which it is possible to extract the piezoresistive anisotropy.

\section{piezoresistive anisotropy of random resistor networks}
\label{piezo}
In this section, we describe in full generality how to extract from a
random resistor network model the relevant piezoresistive coefficients.
Although real materials like TFRs are quite complex from the point
of view of both microscopic electronic processes and microstructure, we believe
that the essential physics of the piezoresistive response 
is already contained in our simplified model. Possible limitations of our approach
are however discussed in Sec.\ref{concl}.

In common piezoresistive measurements, the sample is subjected to
a geometrical distortion characterized by a strain field with
coefficients $\varepsilon_{ii}$ ($i=x,y,z$) and the piezoresistive
response is obtained by recording the change of resistance (conductance)
under the effect of $\varepsilon_{ii}$. In the following, we
shall consider only situations in which the shearing strains can be neglected,
so that $\varepsilon_{ij}=0$ for $i\neq j$.
In the absence of imposed strains,
we assume that the sample is isotropic and characterized by 
a conductance $G$. Hence, under the effect of $\varepsilon_{ii}$, the
conductance for a voltage drop in the $i$ direction 
changes to $G_i=G+\delta G_i$, where the variation $\delta G_i$
can be expressed in terms of the conductivity change $\delta \sigma_i$
and a geometric factor:\cite{prude3}
\begin{equation}
\label{cond1}
\frac{\delta G_i}{G}=\frac{\delta \sigma_i}{\sigma}-\varepsilon_{ii}+
\varepsilon_{jj}+\varepsilon_{kk},
\end{equation}
where $\sigma$ is the conductivity for the unstrained sample 
and the indexes $i$, $j$, and $k$
assume the values $x$, $y$, and $z$ with cyclic permutations.

We are interested on the intrinsic conductivity change $\delta \sigma_i$
which is governed by the microscopic electronic processes taking place 
in the bulk. In agreement with the empirical relation Eq.(\ref{tunnel1}),
we assume that bulk transport is mainly governed by tunneling processes
between metallic grains and we consider a cubic random-resistor network whose 
unstrained bond conductances 
are either proportional to $\exp(2d/\xi)$, where $d$ is the distance between two
neighbouring sites, or zero. The probability of
having zero bond conductance can be independent of the particular bond considered,
as in bond percolation models, or correlated, as in site percolation models.
The following considerations apply also for more complicated distributions,
like the ones arising in segregated site percolation models where all the
sites within spheres of a given radius centered at random are excluded
from the network,\cite{kusy} or when the non-zero bond conductances have some distribution
of tunneling probability.\cite{grima1,grima2}  

To model the effect of the applied strain, let us assume that the resistor 
network is embedded in a homogeneous elastic medium and that the elastic 
coefficients of the network and the medium are equal. Moreover we assume also that
the directions of the bonds in the network are aligned to the $x$, $y$, and $z$
axes. Therefore, in general, when a strain field $\varepsilon_{ii}$ is
applied on the sample, the conducting bonds change their values according to their
orientation with respect to $\varepsilon_{ii}$. For example, the tunneling distance
for a bond directed along the $i$ axis changes to $d\rightarrow d_i
=d(1+\varepsilon_{ii})$ which, up to the first order in $\varepsilon_{ii}$,
implies a bond conductance change
\begin{equation}
\label{tunnel2}
\exp\left(\frac{2d}{\xi}\right)\rightarrow \exp\left(\frac{2d_i}{\xi}\right)=
\exp\left(\frac{2d}{\xi}\right)\left(1+\frac{2d}{\xi}\varepsilon_{ii}\right). 
\end{equation}
Now if we consider an uniaxial strain
along, for example, the $x$ direction ($\varepsilon_{xx}=\varepsilon$,
$\varepsilon_{yy}=\varepsilon_{zz}=0$), thus the bond conductances directed
along $i=x$ are modified according to Eq.(\ref{tunnel2}), while those directed
along the $y$ and $z$ axes remain unchanged. Therefore, up to first order in $\varepsilon$,
the resulting conductivities can be expressed as 
$\sigma_x=\sigma-\sigma\Gamma_\parallel\varepsilon$,
and $\sigma_y=\sigma_z=\sigma-\sigma\Gamma_\perp\varepsilon$, where we have defined
\begin{eqnarray}
\label{gammapar}
\Gamma_\parallel &=&-\frac{\delta \sigma_x}{\varepsilon \sigma}, \\
\label{gammaper}
\Gamma_\perp &=&-\frac{\delta \sigma_y}{\varepsilon \sigma}=
-\frac{\delta \sigma_z}{\varepsilon \sigma},
\end{eqnarray}
as the longitudinal and transverse piezoresistive coefficients, respectively.\cite{grima2}
The above reasoning holds true also for uniaxial strains along the $y$ and the $z$ axis and,
since the problem is linear, for a general strain field $\varepsilon_{ii}$ ($i=x,y,z$) 
the  conductivity variations $\delta \sigma_i$ reduce to:
\begin{equation}
\label{G2}
\frac{\delta \sigma_i}{\sigma}=-\Gamma_\parallel\varepsilon_{ii}-
\Gamma_\perp(\varepsilon_{jj}+\varepsilon_{kk}). 
\end{equation}
As already pointed out in the introduction, the commonly measured quantities used to
extract the strain sensitivity of the sample are the piezoresistive
gauge factors which for a general strain are defined as:
\begin{equation}
\label{gf}
K_{ij}=-\frac{\delta G_i}{\varepsilon_{jj}G}.
\end{equation}
If we assume that the strains $\varepsilon_{ii}$ are known, then 
Eqs.(\ref{cond1},\ref{G2}) permit to express the different gauge factors $K_{ij}$
in terms only of the two {\it intrinsic} piezoresistive coefficients $\Gamma_\parallel$
and $\Gamma_\perp$. 
For example, in a typical cantilever bar experiment with the cantilever main axis 
directed along the $x$ direction, the strains are approximatively 
$\varepsilon_{xx}=\varepsilon$, $\varepsilon_{yy}=-\nu\varepsilon$,
and $\varepsilon_{zz}=-\nu'\varepsilon$, where $\nu$ and $\nu'$ are the Poisson ratios
of the cantilever and the resistive sample, respectively.\cite{prude3,note}
By using Eqs.(\ref{cond1},\ref{G2}), the longitudinal ($K_{\rm L}$) and 
transverse (K$_{\rm T}$) piezoresistive gauge factors are:
\begin{eqnarray}
\label{KL}
K_{\rm L}\equiv K_{xx}&=&(1+\Gamma_\parallel)+(1-\Gamma_\perp)(\nu+\nu'), \\
\label{KT}
K_{\rm T}\equiv K_{yx}&=&-(1+\Gamma_\parallel)\nu 
-(1-\Gamma_\perp)(1-\nu') .
\end{eqnarray}
From a measurement of $K_{\rm L}$ and $K_{\rm T}$, the above expressions permit
to extract $\Gamma_\parallel$ and $\Gamma_\perp$ which are intrinsic quantities,
{\it i. e.} they do not depend on the particular strain applied.
Note that instead of measuring $K_{\rm L}$ and $K_{\rm T}$, the
piezoresistive coefficients can also be extracted from 
Eqs.(\ref{cond1},\ref{G2}) by measuring a gauge factor
under two different imposed strains. For example, as in 
Ref.\onlinecite{fawcett}, a cantilever bar measurement provides a
first value of $K_{\rm L}$, Eq.(\ref{KL}), and a successive measurement 
of the longitudinal gauge factor under hydrostatic pressure provides
a second value of $K_{\rm L}$. These two values are then sufficient to
extract $\Gamma_\parallel$ and $\Gamma_\perp$. 

The importance of $\Gamma_\parallel$ and $\Gamma_\perp$ resides on the fact that
they permit to extract useful informations on the percolative nature of transport.
To illustrate this point, we show in Fig. 1 the results of numerical Monte Carlo calculations
of $\Gamma_\parallel$ and $\Gamma_\perp$ (filled circles) for a site percolation model in which
a concentration of sites $x$ is removed at random. For each missing site
the bond conductances connecting the six neighbouring sites are set equal to zero, while
the remaining bonds have conductance $\exp(2d/\xi)$ with $2d/\xi=4$ when unstrained.
Note that in this model, the fraction of conducting bonds $p$ is equal to $x^2$,
since a bond is present only if the sites at both ends are present.\cite{kirk}
We calculate the total conductance by solving numerically the Kirchhoff equations
for all the nodes of the cubic network. In practice, we impose a unit voltage
difference between
two opposite sides of the network with periodic boundary conditions to the
remaining sides. The piezoresistive coefficient $\Gamma_\parallel$ ($\Gamma_\perp$) 
is then obtained by calculating, for a fixed configurations of bond resistors, 
the difference in conductance when $\varepsilon=0$ and $\varepsilon=0.01$ for strain
directed parallel (ortogonal) to the imposed voltage drop direction.

When there are no missing sites ($x=1$, $p=1$), the current flows exclusively along
paths directed along the direction of the voltage drop. In this case, 
the longitudinal piezoresistive coefficient
is $\Gamma_\parallel=2d/\xi=4$ while the transverse one $\Gamma_\perp$ is zero 
[see Eqs.(\ref{gammapar},\ref{gammaper})]. 
When sites are removed ($x<1$), $\Gamma_\parallel$
gets reduced and at the same time the transverse coefficient is enhanced in such a
way that $\Gamma_\parallel > \Gamma_\perp$.
This behavior is due to the fact that as sites are removed from the network,
the missing bonds force the current to flow also along directions perpendicular
to the voltage drop. Hence, when the strain is along the voltage drop, the
current visits also regions where instead the bonds are unstrained (reduction of
$\Gamma_\parallel$) while when the voltage drop is perpendicular to the
direction of $\varepsilon$, the current is influenced also by regions where
the bonds are strained (enhancement of $\Gamma_\perp$).
This trend gets amplified as $x$ is further reduced and at
the percolation thresold ($p_c=x_c^2\simeq 0.098$) $\Gamma_\perp\rightarrow \Gamma_\parallel$. 
Qualitatively, this behavior 
is observed also when the bond conductances have some distribution or when different
statistics like that of bond percolation models are considered.\cite{grima1,grima2} 
Hence, as a general rule, the
longitudinal and transverse piezoresistive coefficients become equal as the random
resistor network reaches its percolation thresold. 

In relation to the gauge factors $K_{\rm L}$ and $K_{\rm T}$, it is worth to point out
that, from Eqs.(\ref{KL},\ref{KT}), $K_{\rm L}-K_{\rm T}=(\Gamma_\parallel-\Gamma_\perp+2)(1+\nu)$
in accord with Ref.\onlinecite{prude3}. However, since
$\Gamma_\parallel > \Gamma_\perp$ for $p_c < p \le 1$, it is not justified to
assume $K_{\rm L}-K_{\rm T}=2(1+\nu)$,\cite{prude3} unless the system 
is very close to its percolation thresold. Instead a more general relation is
$K_{\rm L}-K_{\rm T}\ge 2(1+\nu)$, which holds true in the whole range of 
$p\ge p_c$ values.

\begin{figure}[t]
\protect
\centerline{\epsfig{figure=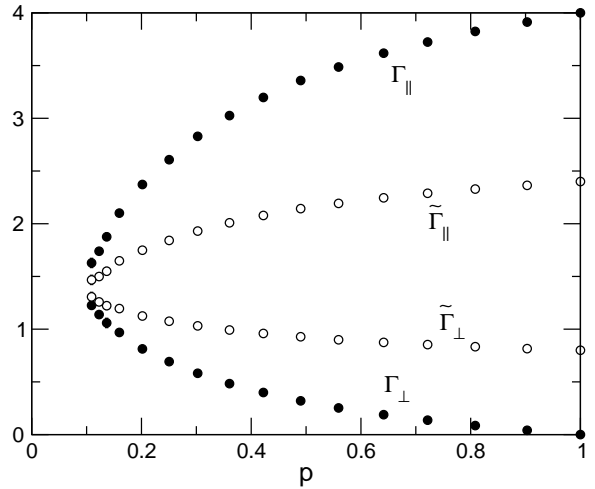,width=20pc,clip=}}
\caption{Monte Carlo calculations of longitudinal ($\Gamma_\parallel$)
and transverse ($\Gamma_\perp$) piezoresistive coefficients of
site percolation cubic resistor networks with number of sites 
up to $40\times 40\times 40$. $p=x^2$ is the fraction of bonds
present for a concentration $x$ of sites. The critical bond fraction
is $p_c=x_c^2\simeq 0.098$.
Filled circles refer to networks with bonds oriented along the
$x$, $y$, and $z$ axes, while the open circles are the effective
longitudinal ($\widetilde{\Gamma}_\parallel$) and transverse
($\widetilde{\Gamma}_\perp$) coefficients resulting from an average over
bond orientations [see Eqs.(\ref{tildea},\ref{tildeb})]. }
\label{fig1}
\end{figure}

Summarizing the results of Fig.1, in the presence of an external uniaxial strain,
the network is anisotropic when $p\sim 1$ while is electrically quasi-isotropic
when $p\sim p_c$. The metal-to-insulator transition can therefore be viewed also as
a piezoresistive anisotropic-to-isotropic transition when $p\rightarrow p_c$. 
We call the quantity which characterizes this transition the piezoresistive
anisotropy factor $\chi$, defined as:
\begin{equation}
\label{chi}
\chi=\frac{\Gamma_\parallel-\Gamma_\perp}{\Gamma_\parallel}.
\end{equation}
As shown in Fig. 2, $\chi$ (filled circles) is equal to the unity for $p=1$ while it 
decreases monotonically as the system moves towards its
percolation thresold and, as shown in log-log plot of the inset of Fig.2, $\chi$
goes to zero by following a
power law behavior in the critical region $(p-p_c)\ll 1$:
\begin{equation}
\label{crit1}
\chi\sim (p-p_c)^\lambda,
\end{equation}
where $\lambda$ is a critical exponent.\cite{grima2} A fit of the numerical
data of Fig.2 to Eq.(\ref{crit1}) leads to $\lambda=0.44\pm 0.07$.
We have tested that Eq.(\ref{crit1}) holds true also for bond percolation models and for
conducting bonds having a distribution of tunneling probability.\cite{grima2}

The power law behavior of Eq.(\ref{crit1}) is directly related to the transport
properties of anisotropic bond percolation models studied some
time ago.\cite{shklo,lobb,sary} In those works, percolating networks with random
bond conductances were defined in such a way that the conducting bonds were equal 
along two directions, for example $y$ and $z$, but different from those along the 
third direction, that is $x$. Topological considerations,\cite{shklo}
renormalization group analysis,\cite{lobb} and numerical calculations\cite{sary}
showed that the quantity $\sigma_y/\sigma_x-1=\alpha(p)$ has a power law behavior 
$\alpha(p)\sim(p-p_c)^{\lambda'}$ close to the percolation thresold, independently
of the microscopic bond anisotropy.\cite{lobb} In our model, bond anisotropy is induced
by the effect of an applied strain, Eq.(\ref{tunnel2}), so that the two situations are
equivalent. It is in fact easy to see from Eqs.(\ref{gammapar},\ref{gammaper}) 
that for our piezoresistive
model $\sigma_y/\sigma_x-1\simeq (\Gamma_\parallel-\Gamma_\perp)\varepsilon \propto \chi$, so that
the two exponents $\lambda$ and $\lambda'$ are equal. Since the exponent $\lambda$
does not depend on the anisotropy of the microscopic bond conductances,
a measurement of the piezoresistive anisotropy factor $\chi$ is in principle a practical
and alternative way to study the role of anisotropy in percolative conductors.

In relation to TFRs, a measurement of the piezoresistive anisotropy factor $\chi$
as a function of metallic volume concentration would be an useful tool to
characterize their transport properties. The advantage of $\chi$ over other
quantities like for example $K_{\rm L}-K_{\rm T}$ is that it measures the
anisotropy without being seriously influenced by other factors like the
change of tunneling distances with the metallic concentration x. By lowering
$x$ in fact the mean distance between metallic granules should increase leading
to an enhancement of the tunneling distance $d$. This situation can be modeled
by assuming that $d$ is some function of the bond probability $p$, $d(p)=d f(p)$, where 
$f(p)$ increases as $p$ is reduced and is independent of strain. 
It is easy to see that, if the unstrained
conducting bonds have conductance $\exp(2d/\xi)=\exp(2df(p)/\xi)$, then the
longitudinal and transverse piezoresistive coefficients get multiplied by $f(p)$.
This factor exactly cancels out in the definition of $\chi$, Eq.(\ref{chi}).

Let us conclude this section by pointing out a relation between conductance
and piezoresisitve response different from Eq.(\ref{empi}).
As already pointed out in the introduction, 
TFRs should be quite close to a metal-insulator transition, since the volume
concentration is small and the conductance $G$ shows a power 
law behavior.\cite{carcia,szpytma,bobran,kusy}
Hence, if TFRs are actually in the critical regime, then as a function of metallic
concentration the piezoresistive anisotropy
factor should follow Eq.(\ref{crit1}) and consequently:
\begin{equation}
\label{crit2}
G\sim \chi^{\lambda/t}=
\left(\frac{\Gamma_\parallel-\Gamma_\perp}{\Gamma_\parallel}\right)^{\lambda/t},
\end{equation}
where $t$ is the critical exponent of the unstrained conductance
$G\sim(p-p_c)^t$.

\begin{figure}[t]
\protect
\centerline{\epsfig{figure=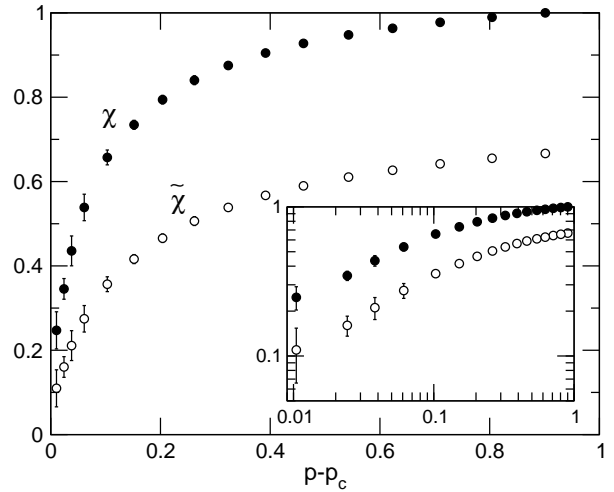,width=20pc,clip=}}
\caption{Piezoresistive anisotropy factor as a function of $p-p_c$ for the 
data of Fig.1. Filled circles ($\chi$) refer to the $x$, $y$, $z$ oriented network
while open circles ($\widetilde{\chi}$) are the results of an average over
bond orientations [see Eq.(\ref{chitilde})]. In the inset, the same data are
plotted in a log-log scale.}
\label{fig2}
\end{figure}

\section{Average over bond orientations}
\label{orientation}
Equation (\ref{G2}) of the last section has been obtained by assuming that
the bond directions are aligned parallel to the $x$, $y$, and $z$ axes
of an ortogonal frame. In this way, under the effect of an applied strain
$\varepsilon_{ii}$, the conducting bonds change according to the simple relation
of Eq.(\ref{tunnel2}). In this section we generalize Eq.(\ref{G2}) in order
to describe networks with bonds oriented at random with respect to the reference
frame. As already pointed out in the introduction, we expect that in this more 
realistic situation
the longitudinal and transverse piezoresistive responses should be different
from those obtained for the oriented network, at least for $p\simeq 1$.

Instead of studying a network with bond directions changing from site to site,
we consider an ensemble of perfectly cubic networks oriented at random. 
The piezoresistive responses will then be obtained by averaging over all
network orientations. We start by choosing
the three bond directions of the cubic network as oriented 
along the $x'$, $y'$, and $z'$ axes which constitute an orthogonal
frame $A'$ rotated by $\hat{R}$ with respect to the reference frame $A$ specified
by the $x$, $y$, and $z$ axes. The rotation matrix $\hat{R}$ is given by the
triple product of successive rotations defined by Euler angles. To fix an explicit
rotation matrix, let us consider a $z$-$x$-$z$ transformation given by a
rotation of angle $\phi$ around the $z$-axis, followed by rotations of
$\theta$ and $\psi$ around the the new $x$ and $z$ axes, respectively.
With this choice of angles, $\theta$ can take values from $0$ to $\pi$,
while $\phi$ and $\psi$ from $0$ to $2\pi$.\cite{arnold}
By using the usual rules of strain transformations, the strains $\varepsilon'_{ij}$
in reference frame $A'$ are given by:\cite{barber}
\begin{equation}
\label{strains}
\varepsilon'_{ij}=\sum_l R_{il}R_{jl}\varepsilon_{ll},
\end{equation}
where $R_{ij}$ are elements of the matrix $\hat{R}$ and, as before, we 
have neglected the shearing strains with respect to frame $A$.
Now, let us consider in the absence of strains a conducting bond directed 
along the $x'$ direction having tunneling length $d_{x'}=d$. Under the effect of 
Eq.(\ref{strains}), the bond length changes to 
\begin{eqnarray}
&&\sqrt{(d+\delta d_{x'})^2+(\delta d_{y'})^2+
(\delta d_{z'})^2} \nonumber \\
&&= d\sqrt{(1+\varepsilon'_{xx})^2+(\varepsilon'_{xy})^2+
(\varepsilon'_{xz})^2} \nonumber \\
&&=d(1+\varepsilon'_{xx})+{\cal O}(\varepsilon'_{xx})+{\cal O}(\varepsilon'_{yy})
+{\cal O}(\varepsilon'_{zz}),
\end{eqnarray}
where $\delta d_{i'}$ is the variation of length along the $i'=x',y',z'$ axis.
Compared to $\varepsilon'_{xx}$, the effect of the shearing strains $\varepsilon'_{xy}$  and 
$\varepsilon'_{xz}$ is therefore of higher order and can be neglected. 
Hence for a bond 
along the $i'$ direction the bond tunneling length changes simply to 
$d\rightarrow d_{i'}=d(1+\varepsilon'_{ii})$. At this point, with respect to the
$A'$ frame, we can follow the same reasonings of last section so that
the conductivity changes are of the same form of Eq.(\ref{G2}):
\begin{eqnarray}
\frac{\delta \sigma'_i}{\sigma}&=&-\Gamma_\parallel\varepsilon'_{ii}-
\Gamma_\perp(\varepsilon'_{jj}+\varepsilon'_{kk}) \nonumber \\
&=&-\Gamma_\perp \Delta+(\Gamma_\perp-\Gamma_\parallel)\varepsilon'_{ii},
\label{G7}
\end{eqnarray}
where $\delta \sigma'_i$ is the variation of conductivity $\sigma'_i$
in the $A'$ frame and we have introduced for later convenience
the volume dilatation $\Delta=\sum_j \varepsilon'_{jj}$.

Now, the local Ohm law is $j'_i=\sigma'_i E'_i$,
where $j'_i$ and $E'_i$ are the components with respect to $A'$
of the current density  and the electric field, respectively. These are related to
the corresponding quantities in frame $A$ by $j'_i=\sum_j R_{ij}j_j$ and 
$E'_i=\sum_j R_{ij}E_j$, so that $j_i=\sigma_{ij}E_j$ where:
\begin{equation}
\label{eq3}
\sigma_{ij}=\sum_l R^{-1}_{il}\sigma'_l R_{lj}.
\end{equation}
Since the dilatation $\Delta$ is invariant under rotation, 
equation (\ref{eq3}) reduces to:
\begin{eqnarray}
\label{eq4}
\sigma_{ij}&=&\delta_{ij}\sigma+R^{-1}_{il}\delta \sigma'_l R_{lj} \nonumber \\
&=&\delta_{ij}\sigma(1-\Gamma_\perp \Delta)+
\sigma(\Gamma_\perp-\Gamma_\parallel)R^{-1}_{il}\varepsilon'_{ll} R_{lj},
\end{eqnarray}
where we have used Eq.(\ref{G7}) and $\delta_{ij}$ is the Kronecker delta symbol.
Finally, the average over the ensemble of networks with different
bond orientations is achieved by performing the average over the
angles $\theta$, $\phi$, and $\psi$ appearing in $\hat{R}$. 
For a general function $f(\theta,\phi,\psi)$, this is achieved by 
\begin{eqnarray}
\langle f\rangle&=&\frac{\int_0^\pi d\theta\int_0^{2\pi}d\phi \int_0^{2\pi} d\psi
J(\hat{R}) f(\theta,\phi,\psi)}
{\int_0^\pi d\theta\int_0^{2\pi}d\phi \int_0^{2\pi} d\psi J(\hat{R})} \nonumber \\
&=&\frac{1}{8\pi^2}\int_0^\pi \!d\theta\int_0^{2\pi}\! d\phi \int_0^{2\pi} \! d\psi
\sin(\theta) f(\theta,\phi,\psi),
\label{inte}
\end{eqnarray}
where $J(\hat{R})=\sin(\theta)/8$ is the Jacobian of $\hat{R}$.
It is easy to see that when $\varepsilon'_{ll}$ in Eq.(\ref{eq4}) are replaced by
Eq.(\ref{strains}),
all the off-diagonal components of $\sigma_{ij}$ vanish under angle
integrations, $\langle \sigma_{ij}\rangle=\delta _{ij}
\langle \sigma_i\rangle$, and that the averaged conductance variations
$\langle \delta \sigma_i\rangle=\langle \sigma_i\rangle-\sigma$ reduce to:
\begin{equation}
\label{orientG}
\frac{\langle\delta \sigma_i\rangle}{\sigma}=
-\Gamma_\perp \Delta+(\Gamma_\perp-\Gamma_\parallel)
\frac{3\varepsilon_{ii}+\varepsilon_{jj}+\varepsilon_{kk}}{5}. 
\end{equation}
By using $\Delta=\sum_j\varepsilon_{jj}$,
this expression can be recast in a form more similar to Eq.(\ref{G2}):
\begin{equation}
\label{eqfin2}
\frac{\langle\delta \sigma_i\rangle}{\sigma}=
-\widetilde{\Gamma}_\parallel\varepsilon_{ii}-
\widetilde{\Gamma}_\perp(\varepsilon_{jj}+\varepsilon_{kk}),
\end{equation}
where we have introduced the effective longitudinal and transverse
piezoresistive coefficients
\begin{eqnarray}
\label{tildea}
\widetilde{\Gamma}_\parallel&=&\frac{3\Gamma_\parallel+2\Gamma_\perp}{5}, \\
\label{tildeb}
\widetilde{\Gamma}_\perp&=&\frac{\Gamma_\parallel+4\Gamma_\perp}{5}.
\end{eqnarray}
If all bonds are present and have all equal conductances $\exp(2d/\xi)$,
then from the above equations we obtain $\widetilde{\Gamma}_\parallel=
3\Gamma_\parallel/5$ and $\widetilde{\Gamma}_\perp=\Gamma_\parallel/5$,
since $\Gamma_\perp=0$ and $\Gamma_\parallel=2d/\xi$. Therefore, as expected,
with respect to a network with bonds oriented along the $x$, $y$, and $z$
axes, the longitudinal response is diminished and the transverse one is enhanced.
On the other hand when the network approaches to its percolation thresold,
$\Gamma_\parallel\rightarrow \Gamma_\perp$, also the effective responses become equal
$\widetilde{\Gamma}_\parallel\rightarrow\widetilde{\Gamma}_\perp$.
This behavior is summarized in Fig. 1 where the open circles refer to
$\widetilde{\Gamma}_\parallel$ and $\widetilde{\Gamma}_\perp$ obtained from
the corresponding coefficients for the oriented network via 
Eqs.(\ref{tildea},\ref{tildeb}).

The effect of averaging over all network orientations is therefore
substantial away from the critical region and unimportant when $p-p_c\ll 1$.
This reflects the universality of the piezoresistive response close the
metal-insulator transition which is more evident in the behavior of the 
effective piezoresistive anisotropy factor
\begin{equation}
\label{chitilde}
\widetilde{\chi}=\frac{\widetilde{\Gamma}_\parallel-\widetilde{\Gamma}_\perp}
{\widetilde{\Gamma}_\parallel}=\frac{2\chi}{3+2(1-\chi)},
\end{equation}
which, according to Eq.(\ref{crit1}), shows the power law behavior 
$\widetilde{\chi}\simeq 2\chi/5 \sim (p-p_c)^\lambda$ for $p-p_c\ll 1$ (open circles
of Fig. 2).

\section{Discussion and conclusions}
\label{concl}

In the previous sections, we have shown how to extract the intrinsic
piezoresistive responses $\Gamma_\parallel$ and $\Gamma_\perp$ of a
granular metal from the knowledge of the gauge factors and the 
values of applied strains. The so-obtained longitudinal and transverse coefficients
permit to define a piezoresistive anisotropy factor $\chi$, Eq.(\ref{chi}),
which follows a power law behavior if the system is sufficiently close
to its percolation thresold. Concerning  the properties of TFRs, a measurement
of $\chi$ as a function of the volume fraction could be a useful tool to
investigate the closeness to a metal-insulator transition.
We have also considered the effect of random bond orientations by employing
an average procedure over an ensemble of networks with different orientation.
This scheme has permitted us to obtain a straightforward and simple generalization
of the results obtained for networks with fixed orientations.
Unfortunately, we are not aware of published data from which it is possible
to extract the piezoresistive coefficients as a function of metallic concentrations,
so our predictions cannot yet be tested. We hope that the present study will
encourage acquisition of data capable of estimating $\chi$ as a function of
metallic concentrations.

Before concluding, it is useful to remind possible limitations of the theory
presented in this work. First of all, all through the previous analysis we have assumed
that the granular metal is elastically homogeneous, so that the strain values
inside the bulk can be considered as independent of the position. However, TFRs
are strongly heterogeneous composites in which quite stiff metallic granules are
embedded in an elastically soft glassy matrix. This gives rise to important
fluctuations in the local values of the strain. Since the microscopic tunneling
processes are affected by the local rather than the macroscopical strain values,
such an elastic heterogeneity can influence the piezoresistive response in an important way.
However, this effect influences the absolute values of $\Gamma_\parallel$
and $\Gamma_\perp$, but should leave the piezoresistive anisotropy factor $\chi$
relatively unaffected. This is indeed confirmed by the analysis of a simplified 
model of elastic heterogeneity in TFRs,\cite{grima3} from which actually it
results that $(\Gamma_\parallel-\Gamma_\perp)/\Gamma_\parallel$ is independent
of the metal and glass elastic coefficients. This is of course only a preliminary result
which needs further investigations.

Another basic assumption of our analysis has been the use of a simple tunneling
form for the intergrain hopping. For TFRs, which show a quite weak temperature
dependence of resistivity, this should not be a too serious approximation, and
simple tunneling captures the essential physics, at least as regards piezoresistance
effects. For other granular metals, or for TFRs at very low temperatures,
other processes than intergrain tunneling (grain charging effects, Coulomb gap, etc.)
play an active role in transport. However, as a function of metallic concentration,
we do not expect qualitative deviations from the power law behavior of $\chi$ in the
critical region. Instead, for fixed metallic concentrations, $\chi$ should develop
a temperature dependence proportional to the importance of the other-than-tunneling
contributions. For example, in the statistical description of the variable-range-hopping
mechanism of Mott,\cite{mott} the temperature governs the availability of sites
energetically favourable for hopping.\cite{ambe} This is a kind of correlated
bond percolation model,\cite{kirk} for which the considerations here presented 
apply in a straightforward way.


\begin{references}
\bibitem{note}
Actually, when the cantilever is very thin, the component
$\varepsilon_{yy}$ can safely be neglected.
\end{references}
\end{document}